\documentclass[aps,prl,reprint,superscriptaddress]{revtex4-2}

\usepackage{graphicx} 
\usepackage{amsmath}
\usepackage{amssymb}
\usepackage{comment}
\usepackage{mathtools}
\usepackage{bm}
\usepackage{tikz}
\usetikzlibrary{arrows.meta}

\DeclareMathOperator{\diag}{diag}
\usepackage{amsthm}

\usepackage{hyperref}
\usepackage{babel}

\makeatletter
\let\ORIbbl@fixname\bbl@fixname
\def\bbl@fixname#1{%
  \@ifundefined{languagealias@\expandafter\string#1}
    {\ORIbbl@fixname#1}
    {\edef\languagename{\@nameuse{languagealias@#1}}}%
}
\newcommand{\definelanguagealias}[2]{%
  \@namedef{languagealias@#1}{#2}%
}
\makeatother

\definelanguagealias{en}{english}
\definelanguagealias{eng}{english}

\begin{document}


\title{A Unified Theory of Edge Weights: Stability of General Laplacian Networks from Matrix Phases and Asymmetry Rayleigh Ratios}






\author{Nina Kastendiek}
\affiliation{These authors have contributed equally}
\affiliation{Potsdam Institute for Climate Impact Research (PIK), 
Member of the Leibniz Association, P.O. Box 60 12 03, D-14412 Potsdam, Germany}
\affiliation{Technical University of Munich, Munich Climate Center, 
TUM School of Engineering and Design, Department of Aerospace and Geodesy, 
Earth System Modelling Group}

\author{Jakob Niehues}
\affiliation{These authors have contributed equally}
\affiliation{Potsdam Institute for Climate Impact Research (PIK), 
Member of the Leibniz Association, P.O. Box 60 12 03, D-14412 Potsdam, Germany}
\affiliation{Institut f\"ur Mathematik, Technische Universit\"at Berlin, 
ER 3-2, Hardenbergstrasse 36a, 10623 Berlin, Germany}

\author{Frank Hellmann}
\affiliation{Potsdam Institute for Climate Impact Research (PIK), 
Member of the Leibniz Association, P.O. Box 60 12 03, D-14412 Potsdam, Germany}



\date{\today}

\begin{abstract}
We study the properties and stability of networks with arbitrary Laplacian coupling. Classic approaches to studying networked systems require unrealistic assumptions, including homogeneous node dynamics, one-dimensional and undirected edges, or constant edge weights. We develop a unified formulation of Laplacian-style couplings that drops these assumptions, providing a unified notion for the edge weights of adaptive, directed, and multi-dimensional edges. We show that the recently developed theory of matrix phases can capture essential stability properties of the network and its edges. We quantify the impact of the asymmetry of the higher-dimensional edge dynamics on the system's phase properties by introducing the Asymmetry Rayleigh Ratio. These theoretical advances allow us to derive new sufficient stability conditions for AC power grids, directed diffusion, and the Kuramoto-Sakaguchi model. The resulting conditions are less conservative than the specific results known for these systems.
\end{abstract}


\maketitle

\paragraph{Introduction---}

Mathematically understanding the properties and stability of networked systems is a central challenge in many areas of theoretical physics and beyond \cite{boccalettiComplexNetworksStructure2006, arenasSynchronizationComplexNetworks2008, barratDynamicalProcessesComplex2008}. Classical approaches to network stability, such as the master stability function (MSF), 
crucially rely on strong assumptions, requiring highly simplified models. They 
can not be applied directly to many real-world settings that feature adaptive edge weights, higher-dimensional interactions on the edge, or complex and heterogeneous nodal dynamics.

The central strategy underlying the MSF approach is to decompose the system into local dynamics and network interactions. Under the assumption that all nodes have identical local dynamics, the MSF reduces the stability properties of the networked system to a single function that is evaluated on the spectrum of the coupling matrix \cite{pecoraMasterStabilityFunctions1998,segelApplicationNonlinearStability1976, bernerDesynchronizationTransitionsAdaptive2021}.
For systems with one-dimensional local node dynamics, diagonal dominance of the Jacobian (via Gershgorin's circle theorem \cite{hornMatrixAnalysis2012}) offers simple sufficient conditions for the stability of the networked system. Further, diagonal dominance can be evaluated node by node and does not require evaluating spectral properties. In the context of couplings mediated by weighted graph Laplacians, positiveness of the edge weights implies 
that all eigenvalues of the Laplacian have non-negative real parts.

Our recent results based on robust control theory and the recently developed theory of matrix phases \cite{wangPhasesComplexMatrix2020,chenPhaseTheoryMultiInput2024,zhaoWhenSmallGain2022}, for the first time, provide an approach to studying properties of networked systems with heterogeneous nodes and heterogeneous adaptive edge dynamics in a localized manner \cite{kastendiekPhaseGainStability2025}. 

In this paper, we introduce and study fully general Laplacian-style couplings, without any simplifying assumptions. We introduce generalized edge weights and a generalized incidence matrix that provide a unified description of undirected,
directed \cite{wuAlgebraicConnectivityDirected2005},
adaptive \cite{grossAdaptiveCoevolutionaryNetworks2008, bernerAdaptiveDynamicalNetworks2023},
matrix-weighted \cite{tianMatrixWeightedNetworksModeling2025},
and higher-order couplings \cite{battistonNetworksPairwiseInteractions2020, bickWhatAreHigherOrder2023}.
This approach cleanly separates the topological coupling structure from the dynamical properties and is suitable for stability analysis based on matrix phases.

The matrix phases of the network coupling capture central stability-relevant properties of the network interactions \cite{kastendiekPhaseGainStability2025} and can be bounded by the phases of the generalized edge weights. In undirected Laplacian-style couplings, the generalized edge weights factorize, and the phase analysis is immediate. To understand the impact of edge directedness on the phase of the network coupling, we introduce the Asymmetry Rayleigh Ratio and related quantities, which quantify how strongly the directedness can perturb the phases of the network coupling.

We demonstrate the broad applicability of our approach by deriving sufficient stability conditions for settings that are beyond the reach of many of the classical methods. We first consider AC power grids with line dynamics, where the coupling on the edges is two-dimensional and adaptive, given by currents satisfying their own differential equations. We then consider diffusion on asymmetric and directed networks and the Kuramoto-Sakaguchi model. In all cases, we obtain sufficient stability conditions beyond the state of the art.

\paragraph{A unification of coupling structures---}

We consider general dynamics on a graph with nodes $n$, edges $e$:
\begin{equation}
\label{eq:adaptive network}
\begin{aligned}
    \dot {\bm x}_n &= \bm f_n(\bm x_n) + \bm g_n(\bm x_n, \{\bm x\}_n, \{\bm z\}_n),\\
    \dot {\bm z}_e &= \bm f_e(\bm z_e, \bm x_e).
\end{aligned}
\end{equation}
Here, we denote node and edge states as $\bm x_n$ and $\bm z_{e}$ respectively, and write $\bm x_e$ for the stacked $\bm x_n$ of nodes that are incident to edge $e$, $\{\bm x\}_n$ for the labeled set of states of neighbors of node $n$, and $\{\bm z\}_n$ for the labeled set of states of edges incident on $n$.

The linearized dynamics around a fixed point is
\begin{align}
    \dot{\bm x}_n &= \bm f_{n;n} \bm x_n + \sum_{e \ni n} \bm g_{n;m} \bm x_n + \sum_{e \ni n}  \bm g_{n;e} \bm z_e \, ,\\
    \dot{\bm z}_e &= \bm f_{e;e} \bm z_e + \sum_{m \in e} \bm f_{e;m} \bm x_m \, ,
\end{align}
where $\bm g_{n;m} = \partial_{\bm x_m} \bm g_n$ (evaluated at the fixed point) etc.

We introduce the \textit{stacking matrix} $\bm B_{ne}$ such that $\bm x_n = \bm B_{ne} \bm x_e$ and, conversely, $\bm x_e = \sum_{n \in e} \bm B_{ne}^\top \bm x_n$. The generalized edge weight $\bm W_e(s)$ is defined by
\begin{align}
\bm B_{ne} \bm W_e(s) \bm B_{me}^\top &:= \bm g_{n;m} + \bm g_{n;e} (s - \bm f_{e;e})^{-1} \bm f_{e;m}\, .
\end{align}

In the frequency domain, the Laplace transform of the linear system can then be written in terms of the node-local \textit{transfers} $\bm T_n := (s - \bm f_{n;n})^{-1}$ and the $\bm W_e(s)$:
\begin{align}
\label{eq:adaptive network as transfers: nodes}
    \bm x_n(s) &= \bm T_n(s) \sum_e \bm B_{ne} \bm o_e(s)\, ,\\
\label{eq:adaptive network as transfers: edges}
    \bm o_e(s) &= \bm W_e(s) \bm x_e(s) \, .
\end{align}

Introduce the \textit{incidence projector} $\bm B^\top_e \bm x = \bm x_e$, and the full \textit{generalized incidence matrix} $\bm B^\top$ obtained by stacking the $\bm B^\top_e$, then we can succinctly write:
\begin{align}
\label{eq:adaptive network as transfers: nodes global}
    \bm x(s) &= \bigoplus_n \bm T_n \bm o(s)= \bm T(s) \bm o(s)\, ,\\
\label{eq:adaptive network as transfers: edges global}
    \bm o(s) &= \bm B \bigoplus_e \bm W_e(s) \bm B^\top \bm x(s) = \bm L(s) \bm x(s)\,.
\end{align}

The global \textit{coupling matrix} $\bm L(s)$ and its \textit{edge weight matrices} $\bm W_e(s)$ provide a unified formulation of network couplings, including adaptive, directed, matrix-weighted, and higher-order.
In this formulation, structural assumptions on the coupling on the edges are entirely encoded in the $\bm W_e$. For example, for scalar nodes,
\begin{align}
\bm W^\text{sym}_e = \left[\begin{smallmatrix} 1 & -1\\ -1 & 1 \end{smallmatrix}\right], \;
\bm W^\text{dir}_e = \left[\begin{smallmatrix} 1 & -1\\ 0 & 0 \end{smallmatrix}\right], \;
\bm W^\text{adj}_e = \left[\begin{smallmatrix} 0 & 1\\ 1 & 0 \end{smallmatrix}\right]
\end{align}
define an unweighted undirected Laplacian edge, an unweighted directed Laplacian edge, and the interaction obtained from the adjacency matrix, respectively.

Many established and proposed interaction mechanisms are naturally expressed by imposing a certain structure on $\bm W_e$. This includes the discourse sheafs of \cite{hansenOpinionDynamicsDiscourse2020, bodnarNeuralSheafDiffusion2022} and the matrix-weighted models of \cite{tianMatrixWeightedNetworksModeling2025,suConsensusDirectedMatrixWeighted2023}.


\paragraph{Linear stability from phase analysis---}

The central utility of this formulation of coupling structures is to enable the analysis of the stability properties of coupled systems using tools from matrix phase theory \cite{wangPhasesComplexMatrix2020, chenPhaseTheoryMultiInput2024}. We briefly recall the central idea here. 

\begin{figure}
\centering
\includegraphics[width=0.6\linewidth]{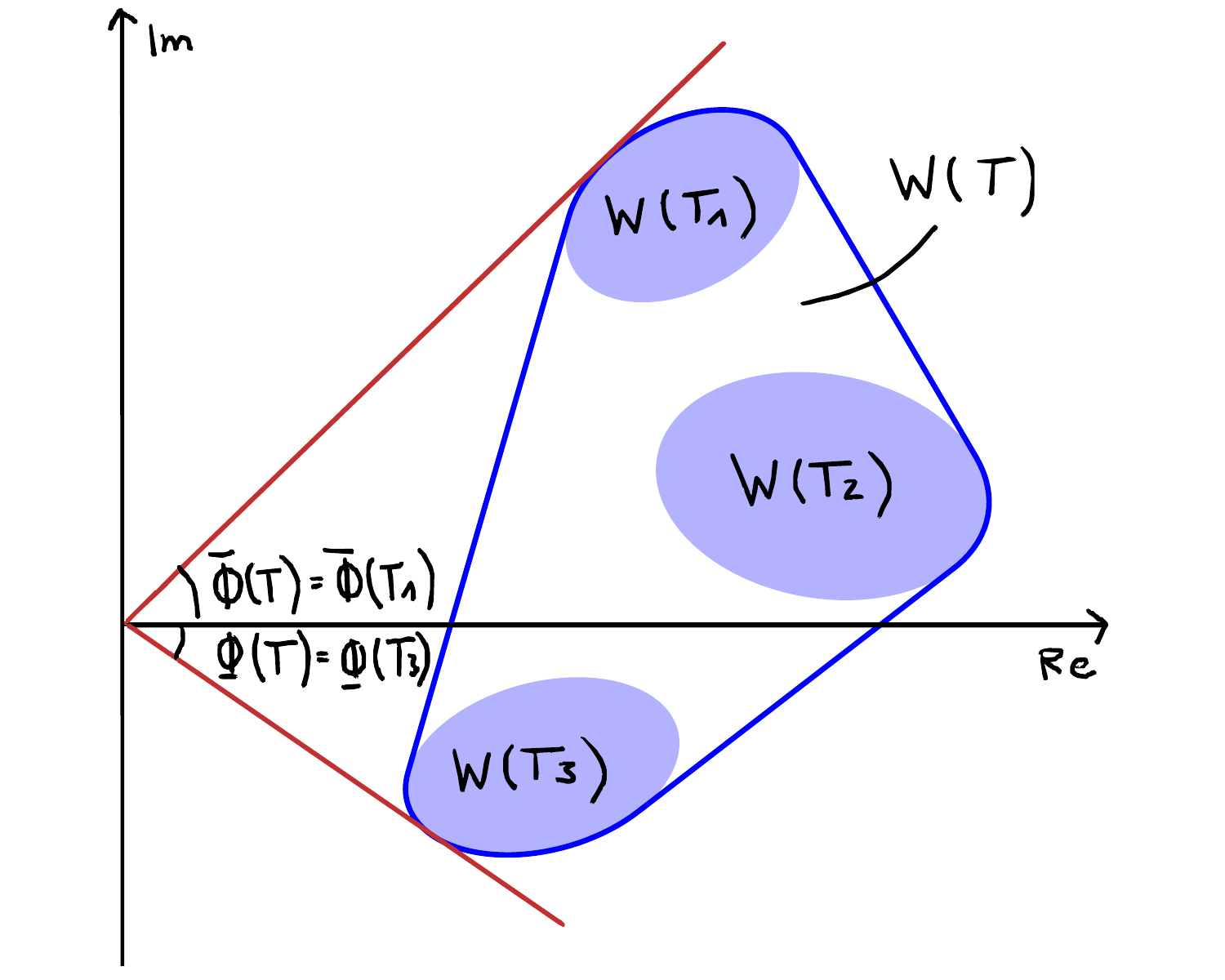}
\caption{The numerical range $W(\bm T) = W\left(\bigoplus_{n=1}^{3} \bm T_n\right)$ is the convex hull of the union of the individual numerical ranges $\bm T_n$.}
\label{fig:numrange}
\end{figure}

If the node-wise $\bm T_n(s)$ and edge-wise $\bm W_e(s)$ represent (semi-)stable dynamical systems, then the stability of the coupled system can be guaranteed by considering the numerical ranges of the nodal $\bm T(s)$ and $\bm L(s)$ on the network side.
The numerical range $W(\bm M)$ of a complex matrix $\bm M$ is defined as
\begin{align}\label{eq:definition numerical range}
W(\bm M) &:= \{\bm v^\dagger \bm M \bm v | \bm v^\dagger \bm v = 1 \}.
\end{align}
If $0\notin W(\bm M)$, $\bm M$ is \textit{sectorial}; if 0 is not in the interior of $W(\bm M)$, it is \textit{semi-sectorial}. The minimum and maximum phases, $\underline \phi (\bm M)$ and $\overline \phi (\bm M)$, are those bounding the smallest sector in which $W(\bm M)$ is contained \cite{chenPhaseTheoryMultiInput2024}. Further, $\overline \sigma (\bm M)$ denotes the largest singular value. Fig.~\ref{fig:numrange} illustrates the numerical range of the nodal transfer $\bm T(s)$.

Under some technical conditions which are recalled in the appendix, it can then be shown that the interconnected system of stable, sectorial $\bm T(s)$ and semi-stable, semi-sectorial $\bm L(s)$ is stable, if 
\begin{align}
    -\pi <\phi(\bm T(s)) + \phi(\bm L(s))&<\pi &\text{for}\, s\in i[0,\omega_c),\\
    \overline\sigma(\bm T(s)) \overline\sigma(\bm L(s))&< 1 &\text{for}\, s\in i[\omega_c,\infty],
\end{align}
where $\phi \in\{\underline\phi,\overline\phi\}$, and $\omega_c>0$ is the \textit{cutoff frequency}, and $i$ is the imaginary unit.
The above phase and gain conditions effectively bound the eigenvalues of the coupled system.

This approach has been used to design synchronizing controllers for heterogeneous agents \cite{wangSynchronizationDiverseAgents2024, wangFirstFiveYears2024}. Our papers \cite{kastendiekPhaseGainStability2025, niehuesSmallSignalStabilityPower2026} show that these tools give excellent results for adaptive networks and, in the power grid case, for edges that can be made Hermitian. In the following, we apply these methods in a much more general setting provided by our generalized edge weights.

\paragraph{Laplacian coupling---}

Our focus in the following will be on Laplacian-type couplings, which we define as couplings where the edge weights of the linearized system have the form
\begin{align}
    \bm W_e(s) = \begin{bmatrix} \bm M_{nm}(s) & - \bm M_{nm}(s)\\ - \bm M_{mn}(s) & \bm M_{mn}(s) \end{bmatrix}.
\end{align}

Note that this definition differs from the matrix-weighted Laplacians considered in \cite{tianMatrixWeightedNetworksModeling2025} in the block-diagonal terms and generalizes the matrix-weighted framework of \cite{suConsensusDirectedMatrixWeighted2023} and the sheaf Laplacian in \cite{hansenOpinionDynamicsDiscourse2020, bodnarNeuralSheafDiffusion2022}.

\paragraph{The undirected case---Phases from edge weights---}

We first demonstrate how phase-based analysis naturally covers the undirected case $\bm M_{nm}(s) = \bm M_{mn}(s)$. In this setting, we can write:
\begin{align}\label{eq:matrix weighted W}
    \bm W_e(s) =  \begin{bmatrix} 1 & -1\\ -1 & 1 \end{bmatrix}\otimes \bm M_e(s).
\end{align}

If the $\bm W_e(s)$ are semi-sectorial, it follows from \eqref{eq:definition numerical range} that the phases of $\bm L$ are bounded by those of the $\bm W_e$. Condition \eqref{eq:phase_cond} can then be broken down into fully node- and edge-local conditions \cite{kastendiekPhaseGainStability2025,niehuesSmallSignalStabilityPower2026}. Further, $\bm W_e(s)$ is semi-sectorial if and only if $\bm M_e(s)$ is semi-sectorial. In this case, we also have $\phi(\bm W_e(s)) = \phi(\bm M_e(s))$.

In particular, the system \eqref{eq:adaptive network as transfers: nodes}--\eqref{eq:adaptive network as transfers: edges} of $\bm T_n$ and $\bm W_e$ is stable if:
(i) Each node transfer $\bm T_n(s)$ is stable and sectorial on $s=i\omega$ with $\max_n \overline\phi (\bm T_n(i\omega)) - \min_n \underline\phi (\bm T_n(i\omega)) < \pi$,
(ii) each edge weight $\bm M_e(s)$ is semi-stable and semi-sectorial on $i\omega$ with $\max_e \overline\phi (\bm M_e(i\omega))- \min_e \underline\phi (\bm M_e(i\omega)) \le \pi$, and
(iii)
for all $n,e$ and $\phi \in\{\underline\phi,\overline\phi\}$:
\begin{align}\label{eq:phase_cond}
    -\pi <\phi(\bm T_n(s)) + \phi(\bm M_e(s))&<\pi &\text{for}\, s\in i[0,\omega_c),\\
    \label{eq:gain_cond}
    \overline\sigma(\bm B\bm B^\top)\overline\sigma(\bm T_n(s)) \overline\sigma(\bm M_e(s))&< 1 &\text{for}\, s\in i[\omega_c,\infty].
\end{align}
Conditions (i)--(ii) ensure that the blocks share a (tilted) half-plane, cf. Fig.~\ref{fig:numrange}.
Note that we do not assume that the undirected Laplacian defined from these $\bm M_e(s)$ is Hermitian. In adaptive dynamical systems, this will typically not be the case.

\paragraph{The directed case---Phases from Asymmetry Rayleigh Ratios---}

In the case of asymmetric coupling, neither the individual $\bm W_e(s)$ nor the full network coupling $\bm L(s)$ is typically semi-sectorial because of asymmetric zero modes.
We address this problem by working orthogonal to the right zero mode and propose the Asymmetry Rayleigh Ratio to bound the phases.

To illustrate the zero mode problems, consider scalar edge weights.
The left and right zero eigenvectors of $\bm L$ coincide exactly if the graph is balanced, i.e., if every node has equal weighted in- and out-degree. If the graph is unbalanced and the eigenvectors do not coincide, general results show that $0$ is in the interior of the numerical range (Theorem 1.6.6 of \cite{hornTopicsMatrixAnalysis1991}), and $\bm L$ is not (semi-)sectorial.

In the context of dynamical systems, we are often interested in the stability in the phase space \textit{orthogonal} to $\ker (\bm L)$. In the following, we will restrict to $\bm v \perp  \ker(\bm L)$, and denote the restricted numerical range $W^\perp(\bm L)$.
This precludes a purely edge-wise phase analysis, but still allows for quasi-local stability conditions determined by local edge properties and some aggregated information of the global network structure.

We begin by decomposing the weights $\bm W_e$ into an \textit{undirected} and a \textit{superdirected} part:
\begin{align}\label{eq:decomposition}
    \bm W_e(s) = \underbrace{\begin{bmatrix} 1 & -1\\ -1 & 1 \end{bmatrix} \otimes \bm M^u_e (s)}_{\displaystyle :=\bm L^u_e(s)} + \underbrace{\begin{bmatrix} 1 & -1\\ 1 & -1 \end{bmatrix} \otimes \bm M^s_e (s)}_{\displaystyle :=\bm L^s_e(s)},
\end{align}
where $\bm M^u_e = \frac{1}{2}(\bm M_{nm} + \bm M_{mn})$ and $\bm M^s_e = \frac{1}{2}(\bm M_{nm} - \bm M_{mn})$.
From now on, we will suppress the $(s)$ for brevity.
Note that neither $\bm L_e^u$ nor $\bm L^s_e$ are necessarily Hermitian.

The local decomposition induces the global split $\bm L = \bm U + \bm S$,
where $\bm U := \bm B \bigoplus_e \bm L^u_e \bm B^\top$ is an undirected matrix-weighted Laplacian, and $\bm S :=\bm B \bigoplus_e \bm L^s_e \bm B^\top$ can be regarded as the skew-Laplacian of a matrix-weighted directed graph \cite{caiNewSkewLaplacian2013}. In what follows, we will always assume that $\ker (\bm L) = \ker (\bm U) = \ker (\bm S)$.

We require that $\bm U$ is sectorial and $\bm S$ acts as a perturbation. Take $\bm v \in \ker(\bm L)^\perp$. To bound the numerical range $W^\perp(\bm L)$ in terms of $\bm S$ and $\bm U$, we introduce the \textit{Asymmetry Rayleigh Ratio} (ARR):
\begin{align}
R(\bm S, \bm U; \bm v) &:= \frac{\bm v^\dagger \bm S \bm v}{\bm v^\dagger \bm U \bm v}\quad \text{such that}\\
\bm v^\dagger \bm L \bm v &= \bm v^\dagger \bm U \bm v \, (1 + R(\bm S, \bm U; \bm v)).
\end{align}
The coefficient $1 + R(\bm S, \bm U; \bm v)$ determines the deformation of the numerical range of $\bm U$ by $\bm S$.
In particular, if $\bm U$ is sectorial, then $\bm L$ is sectorial if and only if $R\neq -1 $ $\forall  \bm v$.

Typically, it will not be possible to evaluate the ARR exactly.
In the following, we derive sufficient estimates.
Detailed calculations are contained in the appendix.

By bounding the ARR we can bound the possible deformations and ensure sectoriality in the space under study. To this end, we introduce the \textit{Asymmetry Rayleigh Coefficient} (ARC):
\begin{align}\label{eq:rho}
    \rho := \max_v |R(\bm S, \bm U; v)|,
\end{align}

If $\bm U$ is sectorial, 
$\rho < 1$ 
guarantees that $\bm L$ is sectorial. Further, $1+R(\bm S,\bm U;\bm v)$ lies in the disk of radius $\rho$ centered at 1 and can rotate $\bm v^\dagger \bm U \bm v$ by at most $\arcsin(\rho)$, so that $\overline \phi(\bm L) \leq \overline \phi(\bm U) + \arcsin(\rho)$ and $\underline \phi(\bm L) \geq \underline \phi(\bm U) - \arcsin(\rho)$. 

Effective stability results can be obtained by bounding $\rho$ from above, e.g.
\begin{align}
\label{eq:rho1}
    \rho \leq \rho_1 &:= \frac{ \max_n \sum_{e \ni n} \alpha_e}{\lambda_2(\bm L^{0})\beta_\text{min} }\quad \text{and}
\\
\label{eq:rho2}
    \rho \leq \rho_2 &:= \sqrt{\frac{\mu_\text{max} \beta_\text{max}}{ \lambda_2(\hat {\bm L}^{0}) \beta_\text{min}^2}},
\end{align}
where $\beta_e := \tfrac12 \lambda_{\min} \left(e^{-i\gamma}\bm M_e^u+e^{i\gamma}(\bm M_e^u)^\dagger\right)>0$ with $\gamma:= (\max_e \overline \phi(\bm M^u_e) + \min_e \underline \phi(\bm M^u_e))/2$,
$\alpha_e := \overline\sigma(\bm M^s_e)$ and
$\mu_e := \alpha_e^2 \left(\frac{1}{\tilde d_n^{\beta}} +\frac{1}{\tilde d_m^{\beta}}\right)$ with $\tilde d^\beta_n := \frac{\sum_{e \ni n} \beta_e}{d^0_n}$, and the subscripts max/min indicate the maximum/minimum across all edges.
${\bm L}^{0}$ is the unweighted Laplacian of the network topology with diagonal entries $d^0_n$,  and $\hat {\bm L}^{0}$ its normalized version.

If $\bm U$ is sectorial, the sectoriality of $\bm L$ can also be guaranteed by 
\begin{align}
\label{eq:RARC condition}
    \xi := \min_v \Re{(R(\bm S, \bm U; v))} > -1,
\end{align}
where we have defined $\xi$ as the \textit{Real Asymmetry Rayleigh Coefficient} (RARC).

If we assume Hermitian edge weights $\bm M_{nm} = \bm M_{nm}^\dagger$, the RARC can be controlled by bounding $\Re(\bm v^\dagger \bm U \bm v) = \bm v^\dagger \bm U \bm v$ and $\Re(\bm v^\dagger \bm S \bm v)$. Further, $\Re(\bm v^\dagger \bm S \bm v)$ can be related to a matrix-weighted generalization of degree imbalance:
Denoting $\bm D^\text{out}_n = \sum_m \bm M_{nm}$ the out-degree, $\bm D^\text{in}_n = \sum_m \bm M_{mn}$ the in-degree, and $\bm \Delta_n = \bm D^\text{out}_n - \bm D^\text{in}_n$ the generalized degree imbalance, we have
\begin{align}
\Re(\bm v^\dagger \bm S \bm v) = \frac12 \sum_{n} \bm v_n^\dagger \bm \Delta_n \bm v_n \; .
\end{align}
Taking $\bm v^\dagger \bm v = 1$, it follows that $\Re(\bm v^\dagger \bm S \bm v) \ge \tfrac12 \min_n \lambda_{\min}(\bm\Delta_n)$ and $\bm v^\dagger \bm U \bm v\ge \lambda_2(\bm L^0) \min_e \lambda_{\min}(\bm M_e^u)$ if $\bm M_e^u \ge 0$.
Hence, a sufficient condition for the sectoriality of $\bm L$ is
\begin{align}\label{eq:xi_1}
    \xi \ge \xi_1 := \frac{\tfrac 12 \min_n \lambda_{\min}(\bm\Delta_n)}{\lambda_2(\bm L^0) \min_e \lambda_{\min}(\bm M_e^u)} > -1\,,
\end{align} 
which is node/edge local except for the algebraic connectivity of the underlying unweighted network topology. 
This condition generalizes results on the positivity of the algebraic connectivity of directed graphs \cite{wuAlgebraicConnectivityDirected2005} to the setting of Hermitian matrix-weighted edges.

Although the estimates will often be conservative, they cleanly go to zero as $\bm M^s_e$ becomes small. Crucially, they remain tractable for adaptive edges, where $\bm W_e(s)$ and thus the estimates depend on $s$. They disentangle the contributions of the global connectivity (given by $\lambda_2({\bm  L}^0)$), the overall coupling strength, and the degree of asymmetry of the dynamics on the edges.

\paragraph{New Stability Results---}

We now demonstrate that the direct phase analysis of the undirected case and the estimates for the ARR can provide novel stability results in paradigmatic models. We first consider AC power grids with line dynamics, then diffusion on asymmetric graphs, and finally the Kuramoto-Sakaguchi model with inertia.

\paragraph{AC power grids---}
\label{sec:power grids}

AC power grids provide a concrete example of undirected, matrix-weighted, and adaptive Laplacian coupling.
The complex-valued current
$j_e$ on an $R$-$L$
transmission line $e=(n,m)$ with terminal voltages $v_n$ and $v_m$ obeys
\begin{align}
    L_e \dot j_e = -Z_e j_e + v_n - v_m\, ,
\end{align}
with inductance $L_e\ge0$,
and impedance $Z_e
= |Z_e| e^{i\zeta_e}$.
Denote $Y_e(s) := (Z_e + s L_e)^{-1}$ the dynamic admittance of edge $e$, and $\overline Y_e(s) := (\overline Z_e + s L_e)^{-1}$.
The matrix weight
\begin{align}\label{eq: transmission line transfer}
    \bm M_e = \begin{bmatrix} g(s) & -b(s)\\ b(s) & g(s) \end{bmatrix}\, ,
\end{align}
with $2 g(s) :=  Y_e(s) + \overline Y_e(s)$ and $2i b(s) :=  Y_e(s) - \overline Y_e(s)$, represents such an edge in Laplace space.

The phases of these weights can be analyzed; detailed calculations are given in the appendix.
Condition \eqref{eq:phase_cond} becomes
\begin{align}
    -\pi < \underline\phi \left(\bm T_n(i\omega)\right)\pm \delta\zeta_e + \arg\left(\vert Z_e\vert - i\omega L_e e^{\mp i\zeta_e}\right),\\
    \pi > \overline\phi\left(\bm T_n(i\omega)\right)\pm \delta\zeta_e + \arg\left(\vert Z_e\vert - i\omega L_e e^{\mp i\zeta_e}\right),
\end{align}
for all $n,e$ and $\omega\in[0,\omega_c)$, and the gain condition \eqref{eq:gain_cond} can be evaluated to
\begin{align}
    \overline\sigma(\bm T_n(i\omega)) < \tfrac{1}{2}d_\text{max}^0 \sqrt{\vert Z_e \vert^2 + L_e^2 \omega^2},
\end{align}
for all $n$, $e$ and $\omega\in[\omega_c,\infty]$,
where $d_\text{max}^0$ is the maximum degree in the network, $\tilde\zeta := \text{mean}_e \zeta_e$ the mean impedance angle, and $\delta\zeta_e := \zeta_e - \tilde\zeta$ the local deviation from the mean.
Note that $\overline\sigma(\bm T_n)$ is expected to vanish for large $\omega$.

A similar analysis was carried out in \cite{huangGainPhaseDecentralized2024}, giving node-local stability conditions with respect to the phase area of the whole network $\bm L$. In contrast, we provide node- and edge-local phases, enabling the edge-local analysis of heterogeneous line impedances for the first time.
We can drop the assumption $\delta \zeta_e = 0$, which has played a crucial role in all prior mathematical work on the stability of power grids that we are aware of. Phase analysis explicitly captures the impact of heterogeneity $\delta \zeta_e$ on stability margins.

\paragraph{Asymmetric diffusion on random graphs---}

Next, we consider first-order Laplacian dynamics on asymmetric networks:
\begin{align}  \label{eq:diffusion}
    \dot {\bm x} = -\bm L \bm x 
    \quad \text{with} \quad \bm W_e = \begin{bmatrix} M_{nm} & -M_{nm}\\ -M_{mn} & M_{mn} \end{bmatrix}\, ,
\end{align}
with real scalar weights $M_{nm} \ne M_{mn}$. Per \eqref{eq:decomposition}, we decompose $\bm W_e$ into undirected and superdirected parts where $M_e^u$ measures the average coupling strength of an edge and $M_e^s$ the asymmetry.

On the global level, $\bm U$ is symmetric and we require $M_e^u > 0$ for all $e$. The symmetric part of $\bm S$ is a diagonal matrix with entries $\Delta_n = D^\text{out}_n - D^\text{in}_n$.
In this setting, condition \eqref{eq:RARC condition} becomes exact. $\bm L$ is sectorial orthogonal to $\ker(\bm L)$ if and only if the RARC $\xi > -1$.

\begin{figure}
\centering
\includegraphics[width=\linewidth]{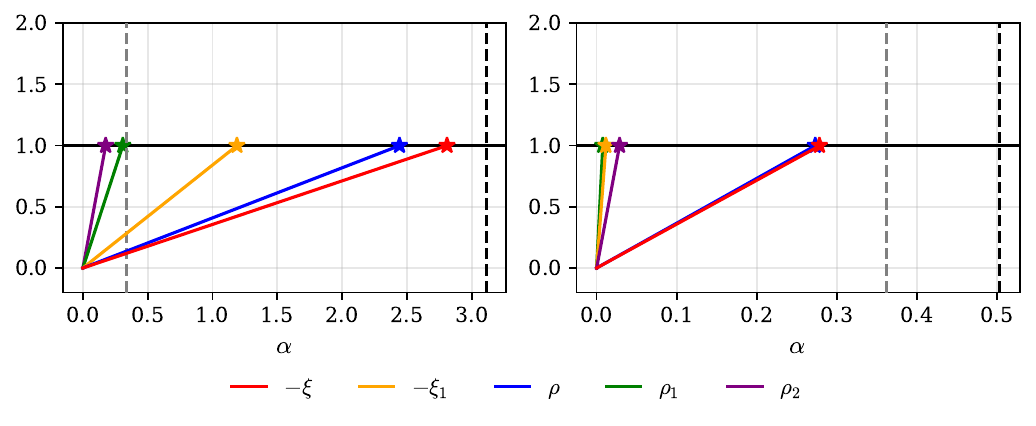}
\caption{Stability bounds for asymmetric Laplacians on random graphs for increasing asymmetry $\alpha$ (left: ER, right: WS). Solid curves show $\xi$, $\xi_1$, $\rho$, $\rho_1$, $\rho_2$. Stars indicate critical $\alpha^\star$, beyond which the corresponding condition in \eqref{eq:rho}–-\eqref{eq:xi_1} fails. Gray dashed line marks loss of diagonal dominance, and black the actual instability threshold.}
\label{fig:simple_Laplacian}
\end{figure}

To evaluate the performance of the RARC estimate \eqref{eq:xi_1} and the conditions based on the ARC \eqref{eq:rho}, we consider Laplacian dynamics on two types of random graphs: an Erdős–Rényi (ER) graph with $N=200$, $p=0.5$, and a Watts–Strogatz (WS) graph with $N=200$, $k=4$, $p=0.1$. We sample $M_e^u$ uniformly from $\left[\frac{1}{2},\frac{3}{2}\right)$ and define $M_e^s=q_e r_e \alpha$, with $q_e$ sampled uniformly from $\left[\frac{1}{2},\frac{3}{2}\right)$ and $r_e$ from $\{-1,+1\}$. We then evaluate the stability conditions for increasing \textit{asymmetry} $\alpha$ (Fig.~\ref{fig:simple_Laplacian}). 

As noted, $-\xi<1$ gives the exact sectoriality condition. The global bound $\rho<1$ is closely aligned with this threshold. In dense ER graphs, the quasi-local bound $-\xi_1<1$ is less conservative than a direct analysis of the Laplacian using diagonal dominance and can prove stability in a regime where some edge weights are negative.
In the WS setting, the bound $\rho_2<1$ is the strongest quasi-local approximation. However, all bounds become more conservative, and sectoriality itself is more restrictive than diagonal dominance.

\paragraph{Kuramoto-Sakaguchi with Inertia---}

We next consider the linear stability of the Kuramoto-Sakaguchi model with inertia
\begin{align}\label{eq:Kuramoto Sakaguchi}
    m_n \ddot x_n + \gamma_n \dot x_n =  - \sum_m  M_{nm} (x_n - x_m),
\end{align}
where $m_n>0$ and $\gamma_n>0$ denote inertia and damping, respectively. The coupling weights are functions of the fixed point around which we linearize, and are asymmetric ($M_{nm}\neq M_{mn}$). We require $\gamma_n/m_n=\mathrm{const}$ for all $n$, so the synchronization subspace $\mathrm{span}\{\bm 1\}$ and its orthogonal complement $\bm 1^\perp$ are invariant, and the dynamics can be restricted to $\bm 1^\perp$.

The system can be represented by the transfers
$\bigoplus_n T_n$ and $\bm L:=\frac{1}{s} \bm B \bigoplus_e \bm W_e \bm B^\top$ with $T_n(s) = \frac{1}{m_ns+\gamma_n}$ and $\bm W_e$ as in \eqref{eq:diffusion}.
Further details are given in the appendix.

As in the previous example, we decompose $\bm W_e$ into undirected and superdirected contributions and get the same conditions for sectoriality. However, sectoriality of $\bm L$ is not sufficient for stability here. Due to the nodal contributions $T_n(s)$, we need to consider the phase sector of $\bm L$ and the analysis purely in terms of the RARR $\xi$ does not apply here. The phase condition \eqref{eq:phase_cond} is valid for all $\omega < \omega_c$ with
\begin{align}
    \omega_c = \frac{\gamma_n}{m_n} \tan \left(\frac{\pi}{2} - \overline \phi(\bm L)\right).
\end{align}
On the subspace orthogonal to $\ker(\bm L)$, we have $\overline{\phi}(\bm L) \leq \arcsin \rho \le \arcsin \rho_{1/2}$.

The system is thus stable if the gain condition  \eqref{eq:gain_cond} is satisfied for all connected pairs of $n$ and $m$ 
\begin{align}
   d^0_{\text{max}} \sqrt{2(M_{nm}^2+M_{mn}^2)} < \omega_c \sqrt{\gamma_n^2 + m_n^2 \omega_c^2}\,.
\end{align}

For the numerical examples, we use the same ER and WS graph realizations as in the previous section. We sample $m_n$ uniformly from $\left[\frac{1}{2},\frac{3}{2}\right)$ and choose 
$\gamma_n/m_n = 2 \sqrt{\text{mean}_n d^{0}_n}$ where $d^{0}_n$ denotes the unweighted degree of node $n$ in the underlying network topology.

\begin{figure}
\centering
\includegraphics[width=\linewidth]{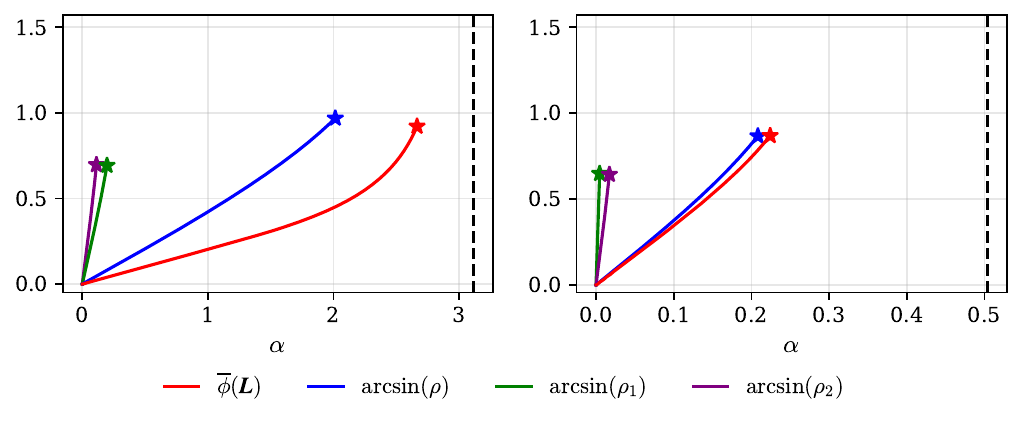}
\caption{Stability bounds for the Kuramoto-Sakaguchi model with inertia on random graphs for increasing asymmetry $\alpha$ (left: ER, right: WS). Solid red curve shows $\overline \phi (\bm L)$, the remaining solid curves are its approximations. Stars indicate the critical $\alpha^\star$, beyond which the mixed phase-gain condition fails. Black dashed line marks the instability threshold.}
\label{fig:second_order}
\end{figure}

Node-local stability conditions for this system were derived in \cite{skarStabilityMultimachinePower1980}: $\gamma_n \ge 2m_n\sum_{m\neq n}M_{nm}$ for all $n$. The damping required is very large, and these conditions are never satisfied for the more realistic setting we investigate here. In contrast, as seen in Fig.~\ref{fig:second_order}, we find that phase-based stability analysis of the full coupling, as well as the ARC $\rho$, works well, although the quasi-local approximations $\rho_1$ and $\rho_2$ are fairly conservative.

\paragraph{Discussion---}

Our results demonstrate that phase-based analysis provides stability conditions for highly general types of Laplacian dynamics. Even in well-studied and established contexts, we obtain novel stability conditions that outperform state-of-the-art methods. These conditions can effectively disentangle local dynamical properties and global structure. They complement the MSF approach by providing sufficient stability conditions in systems where the MSF is not applicable. 
A central result is that the introduced Asymmetry Rayleigh Ratio successfully captures the impact of directedness and asymmetry on the stability of the Laplacian.

In the case of undirected, but heterogeneous, matrix-weighted and adaptive edges studied in the AC Power grid example, we obtained edge-node local conditions, explicitly parametrized by the edge-local heterogeneity.

In the asymmetric examples, we saw that the ARC and RARC 
successfully capture the destabilizing impact of asymmetry. In the dense ER network, our quasi-local conditions based on the degree imbalance were less conservative than diagonal dominance arguments and guaranteed stability in a setting where individual edge weights became negative.

The approach introduced here is highly general. The generalized incidence matrix framework also covers higher-order edges, and it is also possible to decompose the coupling into larger motifs to consider their phases ``at once''. 
Further, the decomposition of edges into undirected and superdirected parts underlying the ARR is not the only natural decomposition.

Finally, we saw that the quasi-local estimates to bound the ARC were typically conservative. We expect that these can be improved by considering weighted Cheeger-type bounds along the lines of \cite{koutisGeneralizedCheegerInequality2023}.

The code used to generate Figures \ref{fig:simple_Laplacian} and \ref{fig:second_order} is available on Zenodo \cite{code}.

\bibliography{zotero, morebib}
\appendix

\section{Local gain and phase stability condition}
For the handling of imaginary-axis poles (semi-stability) and semi-sectorial matrices, we refer to \cite{chenPhaseTheoryMultiInput2024,zhaoWhenSmallGain2022,niehuesSmallSignalStabilityPower2026,kastendiekPhaseGainStability2025}.
In short, one has to make $\varepsilon$-detours around imaginary-axis poles and zeros when evaluating the conditions for $s=i\omega$.
In addition, one has to check that $\bm B\bigoplus_e \bm W_e \bm B^\top$ has constant rank along this indented imaginary axis.
In general, it is assumed that the DC phase center of all matrices is 0 \cite{chenPhaseTheoryMultiInput2024}.
The gain condition can be estimated using $\overline\sigma\left(\bm B\bigoplus_e \bm W_e \bm B^\top\right) \le \overline\sigma(\bm B^\top \bm B) \overline\sigma\left(\bigoplus_e \bm W_e\right) = d_\text{max}^0 \max_e \overline\sigma(\bm W_e)$.

When is it sufficient to use $W^\perp$ for stability analysis?
Consider the system $\bm x = \frac{1}{s}\bm L \bm y$, $ \bm y = -\bm T \bm x$.
If $\bm T$ commutes with the projection onto $\bm 1$, the right zero mode of $\bm L$, then the $\perp$-subsystem drives the $\|$-subsystem unidirectionally.
Thus, the total system is stable orthogonal to $\bm 1$ if the $\perp$-subsystem is stable.
A minimal realization of the $\perp$-subsystem is represented by $W^\perp$.

\section{Estimates for ARR}

In the following, we restrict ourselves to normalized vectors $\bm v^\dagger \bm v=1$ that satisfy $\bm v\perp \ker(\bm L)$.

\paragraph{ARC $\rho_1$ \eqref{eq:rho1}:}
To guarantee $\rho < 1$, we bound $\rho$ from above
\begin{align}
  \rho := \max_v |R(\bm S, \bm U; v)| \le \frac{\max_v |\bm v^\dagger \bm S \bm v|}{\min_v |\bm v^\dagger \bm U \bm v|}.
\end{align}

Suppose $\bm U$ is sectorial and $\max_e \overline \phi(\bm M^u_e)- \min_e \underline \phi(\bm M^u_e)<\pi$ and define the phase center $\gamma:= (\max_e \overline \phi(\bm M^u_e) + \min_e \underline \phi(\bm M^u_e))/2$. Along this common direction $e^{i\gamma}$, each numerical range $W(\bm M_e^u)$ has a positive real projection bounded below by $\beta_e := \tfrac12 \lambda_{\min} \left(e^{-i\gamma}\bm M_e^u+e^{i\gamma}(\bm M_e^u)^\dagger\right)$. Therefore,
\begin{align}
    \min_v |\bm v^\dagger \bm U \bm v| \ge \min_e \beta_e \lambda_2(\bm L^0)
\end{align}

Denote $\alpha_e = \overline\sigma(\bm M^s_e)$. We bound $\max_v |\bm v^\dagger \bm S \bm v|$ as follows: 
\begin{align}
&\left|\bm v^\dagger \bm B \bigoplus\nolimits_e \begin{bmatrix}
        1 &-1\\1 &-1 \end{bmatrix} \otimes \bm M^s_e \bm B^\dagger \bm v \right|\\
    &= \left|\sum_{e=(n,m)} (\bm v_n + \bm v_m)^\dagger \bm M^s_e (\bm v_n - \bm v_m)\right|\\
    &\le \sum_{e=(n,m)} \alpha_e |\bm v_n + \bm v_m||\bm v_n - \bm v_m|\\
    &\le \sum_{e=(n,m)} \alpha_e (|\bm v_n|^2 + |\bm v_m|^2)\\
    &= \sum_n |\bm v_n|^2 \sum_{e \ni n} \alpha_e,
\end{align}
with the AM-GM inequality
$|\bm v_n + \bm v_m||\bm v_n - \bm v_m| \le \frac{|\bm v_n + \bm v_m|^2 + |\bm v_n - \bm v_m|^2}{2} = |\bm v_n|^2 + |\bm v_m|^2$, and
$\bm B_e \bm v =: \begin{bmatrix}
    \bm v_n & \bm v_m
\end{bmatrix}^\top$.
Since $\sum_n |\bm v_n|^2 = 1$,
\begin{align}
    \max_v |\bm v^\dagger \bm S \bm v| \le \max_n \sum_{e \ni n} \alpha_e\, ,
\end{align}
leading to \eqref{eq:rho1}.

\paragraph{ARC $\rho_2$ \eqref{eq:rho2}:}
Define the normalized $\beta$-degree $\tilde d^\beta_n := \frac{\sum_{e \ni n} \beta_e}{d^0_n}$, where $d^0_n$ denotes the unweighted degree. Introduce the edgewise diagonal matrix
\begin{align}
{\bm D}_e^\beta = \begin{bmatrix}
        \tilde d^\beta_n & 0\\0 &\tilde d^\beta_m \end{bmatrix}
\end{align}
such that $\bm B \bigoplus\nolimits_{e} {\bm D}_e^\beta \bm B^\dagger = {\bm D}^\beta = \diag_n (\sum_{e\ni n}\beta_e)$.
Define
\begin{align}
    {\tilde {\bm L}}^s_e &:= {\bm L^s_e}^\dagger \left({\bm D}_e^\beta\right)^{-1} \bm L^s_e \\
    &=\begin{bmatrix}
        1 & 1\\ -1 &-1 \end{bmatrix} \otimes {\bm M^s_e}^\dagger \begin{bmatrix}
        \tilde d^\beta_n & 0\\0 &\tilde d^\beta_m \end{bmatrix}^{-1} \begin{bmatrix}
        1 &-1\\1 &-1 \end{bmatrix} \otimes \bm M^s_e\\
        &=\left(\frac1{\tilde d^\beta_n} + \frac1{\tilde d^\beta_m}\right) \begin{bmatrix}
        1 & -1\\-1 &1 \end{bmatrix} \otimes {\bm M^s_e}^\dagger \bm M^s_e
\end{align}

Now by Cauchy-Schwartz:
\begin{align}
\rho^2 = &\frac{| \bm v^{\dagger} \bm B \bigoplus\nolimits_{e} {\bm D_e^\beta}^{\frac12} {\bm D_e^\beta}^{-\frac12} \bm  L^s_e \bm B^{\dagger} \bm v |^2}
{|\bm v^{\dagger} \bm B \bigoplus\nolimits_{e} \bm L^u_e \bm B^{\dagger} \bm v|^2}\\
 &\leq \frac{| \bm v^{\dagger} \bm B \bigoplus\nolimits_{e} \bm D_e^\beta \bm B^{\dagger} \bm v| | \bm v^{\dagger} \bm B \bigoplus\nolimits_{e} {\bm L^s_e}^\dagger {\bm D_e^\beta}^{-1} \bm L^s_e  \bm B^{\dagger}\bm v |}
     {|\bm v^{\dagger} \bm B\bigoplus\nolimits_{e} \bm L^u_e \bm B^{\dagger} \bm v|^2} \\
     &= \frac{| \bm v^{\dagger} \bm D^\beta \bm v| | \bm v^{\dagger} \bm B \bigoplus\nolimits_{e} \tilde {\bm L}^s_e  \bm B^{\dagger}\bm v |}
     {|\bm v^{\dagger} \bm B \bigoplus\nolimits_{e} \bm L^u_e \bm B^{\dagger} \bm v|^2}
\end{align}
Then we obtain the bound:
\begin{align}
\rho^2 &\leq \max_v \frac{| \bm v^{\dagger} \bm D^\beta \bm v| | \bm v^{\dagger} \tilde {\bm L}^s \bm v |}{|\bm v^{\dagger} \bm L^u \bm v|^2} \\
&\leq \frac{ \beta_\text{max} \mu_\text{max}}{\beta_\text{min}^2} \max_v \frac{| \bm v^{\dagger} \bm D^0 \bm v| | \bm v^{\dagger} \bm L^0\bm v |}{|\bm v^{\dagger} \bm L^0 \bm v|^2}\\
&\leq \frac{ \beta_\text{max} \mu_\text{max}}{\lambda_2(\hat {\bm L^0}) \beta_\text{min}^2}
\end{align}
where $\hat {\bm L^0} =  (\bm D^0)^{-1/2} \bm L^0 (\bm D^0)^{-1/2}$ is the normalized Laplacian of the unweighted graph, and $\mu_e := \alpha_e^2 \left(\frac{1}{\tilde d_n^{\beta}} +\frac{1}{\tilde d_m^{\beta}}\right)$, $\mu_{\text{max}} =\max_e \mu_e$, $\beta_{\text{max}} = \max_e \beta_e$, $\beta_{\text{min}} = \min_e \beta_e$.
Note that there is freedom in choosing the weights in $\hat d^\beta$ to obtain similar results.

\paragraph{RARC $\xi_1$ \eqref{eq:xi_1}:}
We want to ensure $\xi > -1$, assuming Hermitian weights $\bm M_{nm}$ and $\bm M_{mn}$. With $\beta_e>0$ for all $e$, we have $\min_v \Re(\bm v^\dagger \bm U \bm v) = \min_v \bm v^\dagger \bm U \bm v>0$ and $\xi$ can be bounded from below
\begin{align}
\xi = \min_v \Re{(R(\bm S, \bm U; v))} \ge \frac{\min_v \Re(\bm v^\dagger \bm S \bm v)}{\min_v \bm v^\dagger \bm U \bm v}
\end{align}
Since we expect the numerator to be non-positive and the denominator to be positive, we take the minimum of the latter.

Since $\bm M_{nm}$ and $\bm M_{mn}$ are Hermitian, $\Re(\bm v^\dagger \bm S \bm v)$ reduces to a matrix-weighted degree imbalance
\begin{align}
&\Re(\bm v^\dagger \bm S \bm v)\\
     &= \Re \left(\bm v^\dagger \bm B \bigoplus\nolimits_e \left[\begin{smallmatrix}
        1 &-1\\1 &-1 \end{smallmatrix}\right] \otimes \bm M^s_e \bm B^\dagger \bm v\right)\\
    &= \Re \left( \sum_{e=(n,m)} (\bm v_n^\dagger+\bm v_m^\dagger) \bm M^s_e (\bm v_n-\bm v_m) \right) \\
    &= \sum_{e=(n,m)} \left(
    \bm v_n^\dagger\bm M^s_e\bm v_n
    -\bm v_m^\dagger\bm M^s_e\bm v_m
    \right)\\
    &= \sum_n \bm v_n^\dagger \left(
    \sum_{m: e=(n,m)}\bm M^s_e
    +\sum_{m: e=(m,n)}\bm M^s_e
    \right)\bm v_n\\
    &=\frac12 \sum_{n} \bm v_n^\dagger \left (\sum_{m \sim n}(\bm M_{nm} - \bm M_{mn})\right) \bm v_n
\end{align}
Thus, we bound
\begin{align}
    \min_v \Re(\bm v^\dagger \bm S \bm v) \ge \tfrac 12 \min_n \lambda_{\min}(\bm\Delta_n)
\end{align}

\section{Calculations for AC Power Grids}
Denote an edge with fiducial orientation $e=(n,m)$, and the reverse orientation $\overline e = (m,n)$.
\paragraph{Line model}
The impedance is given by $Z_e = R_e + i \Omega L_e$ with resistance $R_e$ and synchronous frequency $\Omega$.
In $dq$-coordinates, the voltages $x_e = \begin{bmatrix}
    v^d_n & v^q_n & v_m^d & v^q_m \end{bmatrix}^\top$ and currents $z_e = \begin{bmatrix}
        j^d_e & j^q_e & j^d_{\overline e} & j^q_{\overline e}
    \end{bmatrix}^\top$ on the edge are related by $z_e = \bm W_e x_e$ with $\bm M_e$ as given in \eqref{eq: transmission line transfer}.
In this case, the phases of $\bm W_e$ and $\bm M_e$ are identical: $\phi(\bm W_e) = \phi(\bm M_e)$, because the Laplacian $\bm L_e^u$ can be factorized.

\paragraph{Node model}
There exist many dynamic node models.
However, all voltage sources are represented as generic $2\times 2$ transfers $\bm T_n'$ that relate nodal currents $j_n = \sum_{e=(n,m)} j_e$ and voltages \cite{niehuesComplexPhaseAnalysis2025, huangGainPhaseDecentralized2024}.

\paragraph{Analysis}
To obtain the least conservative stability conditions, we rotate $\bm M_e$ by the mean impedance angle (cf. \cite{huangGainPhaseDecentralized2024,niehuesSmallSignalStabilityPower2026}).
We therefore analyze the equivalent system 
$\bigoplus_n \bm T_n $ and $ \hat{\bm B} \bigoplus_e \bm M_e' \hat{\bm B}^\top$, where $\bm T_n := \bm T_n' \bm R(-\tilde\zeta)$,
and $\bm M_e' := \bm R(\tilde\zeta) \bm M_e$,
and $\hat{\bm B} := \bm B\bigoplus_e \begin{bmatrix}
    \bm I_2 & -\bm I_2
\end{bmatrix}^\top$, and
 $\bm R(\zeta) := \begin{bmatrix}
        \cos\zeta & -\sin\zeta\\
        \sin\zeta & \cos\zeta
\end{bmatrix}$.
This formulation gives the smallest phase response at small $\omega$.

\paragraph{Phases}
Since $\bm M_e'$ is normal, its phases are given by the arguments of its eigenvalues: $\arg(e^{i\tilde\zeta}Y_e(s)) = -\delta\zeta_e + \arg\left(\vert Z_e\vert - i\omega L_e e^{i\zeta_e}\right)$ and $\arg(e^{-i\tilde\zeta}\overline Y_e(s)) = \delta\zeta_e + \arg\left(\vert Z_e\vert - i\omega L_e e^{-i\zeta_e}\right)$.
The two phases range from $\pm \delta\zeta_e$ at $s=0$ to approaching $-\frac{\pi}{2} \pm \tilde\zeta$ at $s\to i\infty$, respectively, and  coincide at $\omega_e = - r_e t + \sqrt{\Omega^2 + r_e^2 (1+t^2)}$, where $r_e:=R_e/L_e$,  $t:=\frac{1}{2}\left(\tilde r - \frac{1}{\tilde r}\right)$, and $\tilde r$ is the $R$-$L$ ratio of the mean impedance angle: ${\tilde r}^{-1}:= \tan\tilde\zeta$.
Thus, we have $\underline{\overline\phi} (\bm M'_e)=\pm \eta\cdot\delta\zeta_e + \arg\left(\vert Z_e\vert - i\omega L_e e^{\mp \eta i\zeta_e}\right)$ where $\eta:=\text{sgn}(\omega_e - \omega)$. 

The $\bm M_e'$ are semi-stable and their full rank  ensures that $\hat{\bm B} \bigoplus_e \bm M_e' \hat{\bm B}^\top$ has constant rank \cite{kastendiekPhaseGainStability2025}.
It remains to be verified for the node model at hand whether $\bm T_n$ are stable, sectorial, and within a common sector.

\section{Calculations for Kuramoto-Sakaguchi with Inertia}
By introducing the dynamic coupling variable $\kappa_{nm}$, we rewrite Eq.~\eqref{eq:Kuramoto Sakaguchi} as adaptively coupled phase oscillators
\begin{align}
    \dot x_n &= -\frac{\gamma_n}{m_n} x_n + \frac{1}{m_n} \sum_m \kappa_{nm}\\
    \dot \kappa_{nm} &= - M_{nm} (x_n - x_m)
\end{align}

We go to Laplace space and derive the transfers
$\bigoplus_n T_n$ and $\bm L :=\frac{1}{s} \bm B \bigoplus_e \bm W_e \bm B^\top$ with
\begin{align}
    T_n(s) &= \frac{1}{m_ns+\gamma_n}, \qquad \bm W_e  =  \begin{bmatrix} M_{nm} & -M_{nm} \\ -M_{mn} & M_{mn} \end{bmatrix}
\end{align}

For $T_n(s)$ to be a stable transfer, all poles at $s=-\frac{\gamma_n}{m_n}$ must lie in the open left half-plane, which is ensured by all $\gamma_n>0$ and $m_n>0$. 

We set $s=i\omega$ and evaluate phase and gain over $\omega \in [0, \infty]$. For the nodal transfers, we have
$\phi(T_n(i\omega)) = -\tan^{-1}\left(m_n \omega/\gamma_n\right)$ and $\sigma(T_n(i\omega)) = \left(\gamma_n^2 + m_n^2 \omega^2\right)^{-\frac12}$.
Since the edge weights are real, $W(\bm L)$ is symmetric with respect to the real axis, and $\overline{\phi}(\bm L) = -\underline{\phi}(\bm L)$. The largest singular value can be bounded by $\overline \sigma(\bm L) \le \overline\sigma(\bm B \bm B^\top)  \max_e \overline\sigma(\bm W_e) = d^0_\text{max} \max_{e=(n,m)} \sqrt{2(M_{nm}^2 + M_{mn}^2)}$. Multiplication by $\frac{1}{s}$ in $\bm L$ rotates $W(\bm L)$ by $-\pi/2$ and scales it by $\frac{1}{\omega}$.

\end{document}